\documentclass[
superscriptaddress,amssymb,aps,preprint,dvips]{revtex4-1}
\usepackage{amsmath}
\usepackage{graphicx}
\usepackage[dvipdfm,colorlinks=true,linkcolor=blue,citecolor=blue,anchorcolor=blue,urlcolor=blue]{hyperref}
\usepackage{color}

\begin{document}

\title{Frustration induced non-Curie-Weiss paramagnetism in La$_3$Ir$_3$O$_{11}$: a fractional-valence-state iridate}%

\author{J. Yang}
\affiliation{Anhui Province Key Laboratory of Condensed Matter Physics at Extreme Conditions, High Magnetic Field Laboratory, Chinese Academy of Sciences, Hefei 230031, China}
\affiliation{University of Science and Technology of China, Hefei 230026, China}
\author{J. R. Wang}
\affiliation{Anhui Province Key Laboratory of Condensed Matter Physics at Extreme Conditions, High Magnetic Field Laboratory, Chinese Academy of Sciences, Hefei 230031, China}
\author{W. L. Zhen}
\affiliation{Anhui Province Key Laboratory of Condensed Matter Physics at Extreme Conditions, High Magnetic Field Laboratory, Chinese Academy of Sciences, Hefei 230031, China}
\author{L. Ma}
\affiliation{Anhui Province Key Laboratory of Condensed Matter Physics at Extreme Conditions, High Magnetic Field Laboratory, Chinese Academy of Sciences, Hefei 230031, China}
\author{L. S. Ling}
\affiliation{Anhui Province Key Laboratory of Condensed Matter Physics at Extreme Conditions, High Magnetic Field Laboratory, Chinese Academy of Sciences, Hefei 230031, China}
\author{W. Tong}
\affiliation{Anhui Province Key Laboratory of Condensed Matter Physics at Extreme Conditions, High Magnetic Field Laboratory, Chinese Academy of Sciences, Hefei 230031, China}
\author{C. J. Zhang}
\email[]{zhangcj@hmfl.ac.cn}
\affiliation{Institutes of Physical Science and Information Technology, Anhui University, Hefei 230601, China}
\author{L. Pi}
\email[]{pili@ustc.edu.cn}
\affiliation{Anhui Province Key Laboratory of Condensed Matter Physics at Extreme Conditions, High Magnetic Field Laboratory, Chinese Academy of Sciences, Hefei 230031, China}
\affiliation{University of Science and Technology of China, Hefei 230026, China}
\author{W. K. Zhu}
\email[]{wkzhu@hmfl.ac.cn}
\affiliation{Anhui Province Key Laboratory of Condensed Matter Physics at Extreme Conditions, High Magnetic Field Laboratory, Chinese Academy of Sciences, Hefei 230031, China}

\begin{abstract}
Experimental and theoretical studies are performed on La$_3$Ir$_3$O$_{11}$, an iridate hosting a +4.33 fractional valence state for Ir ions and a three-dimensional frustrated structure composed of edge-shared Ir$_2$O$_{10}$ dimers. These features are expected to enhance inter-site hoppings and reduce magnetic moments of Ir ions. However, a spin-orbit driven Mott insulating transport is observed, which is supported by our first principles calculations. Most importantly, geometrical frustration and competing interactions result in a non-Curie-Weiss paramagnetic ground state, revealing no magnetic order down to 2 K. This unusual state is further demonstrated by a theoretical modeling process, suggesting a possible candidate for the spin liquid state.
\end{abstract}

\maketitle

\section{Introduction}

Chemical valence state describes the electron configuration of ions, and always plays an essential role in the physical properties of compounds. In some transition metal oxides (TMOs), the valence state of transition metal ion is not an integer but a fractional number. For the 3$d$ and 4$d$ TMOs, fractional valence state (FVS) is often associated with charge separation or charge order \cite{1,2}, which could further induce magnetoelectric or electron-lattice correlations \cite{3,4}, as a result of the coupled degrees of freedom. It becomes more complex for the 5$d$ TMOs \cite{5,6,7,8}, in which spin-orbit couplings (SOCs) are comparable to electronic correlations and crystal field, giving rise to new quantum states, e.g., the $J_{\rm eff}$=1/2 state in some iridates \cite{5}. The FVS in iridates could be mixed valence states \cite{Doi2004}, or a probabilistically occupied state \cite{9}, which reflects different degrees of electron itinerancy. As a rare representative for the latter condition, La$_3$Ir$_3$O$_{11}$ possesses a +4.33 valence state for Ir ions.  More importantly, such a valence state is a single one for all Ir ions, according to the M\"{o}ssbauer spectroscopy \cite{9}, instead of an average result of mixed +4 and +5 valence states. That is to say, in La$_3$Ir$_3$O$_{11}$ the 5$d$ electrons show both itinerancy and localization, making it difficult to determine the magnetic moment.

From the structural aspect, La$_3$Ir$_3$O$_{11}$ belongs to a unique family, i.e., the KSbO$_3$-type compounds \cite{9,12,13}, whose crystal structure contains Ir$_2$O$_{10}$ dimers composed of edge-shared IrO$_6$ octahedra. The neighboring Ir ions form triangles which are connected in three-dimensional (3D) space. Namely, this crystal structure hosts considerable geometrical frustration \cite{14}, which would further cause strong magnetic frustration. Therefore, the magnetic ground state of La$_3$Ir$_3$O$_{11}$ remains deeply elusive, especially in consideration of the FVS and dimer structure.

In this paper, we report systematic experimental and theoretical studies on La$_3$Ir$_3$O$_{11}$. Structural characterization shows a 3D frustrated Ir network. A spin-orbit driven Mott insulating transport is discovered, which is consistent with our density functional theory (DFT) calculations. Variable-temperature X-ray diffraction (XRD), heat capacity and susceptibility measurements confirm the absence of lattice distortion and magnetic ordering down to 2 K. More importantly, a non-Curie-Weiss paramagnetic ground state is observed, and demonstrated by a theoretical model in terms of competing interactions beyond the simple Curie-Weiss model. The reduced moment is also determined and understood based on the model. This unusual ground state implies a possible quantum spin liquid (QSL) state.

\section{METHODS}

Single crystal La$_3$Ir$_3$O$_{11}$ was grown from high-purity dry La$_2$O$_3$ and IrO$_2$ with a flux of potassium chloride, as well as a small amount of potassium perchlorate. Mixed powders were sealed in an evacuated quartz ampoule, heated to 1050 $^\circ$C in furnace and held for 2 days, then slowly cooled down to 850 $^\circ$C over a period of 100 hours, and finally cooled down with furnace. The crystal structure and phase purity were checked by powder XRD on a Rigaku-TTR3 X-ray diffractometer using Cu K$\alpha$ radiation. Rietveld refinement was performed using the GSAS software package \cite{20,21}. The chemical component characterization was taken on an Oxford Swift 3000 energy dispersive spectrometer (EDS). The electrical and magnetic measurements were taken on a home-built Multi Measurement System (on a Jains-9T magnet) and a Quantum Design MPMS, respectively. The electron spin resonance (ESR) spectra were taken in a BRUKER EMX plus spectrometer with X-band microwave (frequency $\nu$=9.40 GHz).

DFT calculations were performed using the WIEN2k code \cite{22}. The generalized gradient approximation (GGA) of Perdew-Burke-Ernzerhof \cite{23} was employed with a method of GGA+$U$+SO, where $U$ was the effective exchange-correlation potential for Ir 5$d$ orbitals and SO represented the spin-orbit interactions. RK$_{\rm max}$ was set as 7, where K$_{\rm max}$ was the plane wave sector cutoff and R was the minimum LAPW (linearized augmented plane wave) sphere radius; the sphere radii were set as 2.28, 1.98 and 1.62 Bohr for La, Ir and O, respectively. The $k$ mesh consisted of 1000 points.

\section{RESULTS AND DISCUSSION}
\subsection{Structural characterization}

\begin{figure}[htbp]
\center
\includegraphics[width=8.5cm]{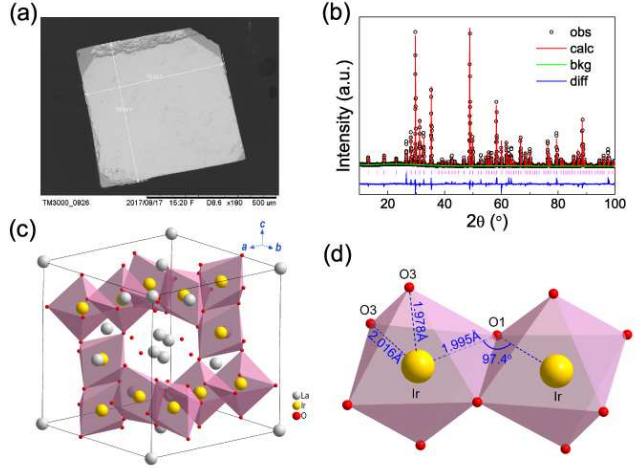}
\caption{(a) SEM image of single crystal La$_3$Ir$_3$O$_{11}$. (b) Powder XRD pattern and Rietveld refinement. Black circles represent the experimental data; red, green and blue curves denote the calculated pattern, background and difference between the experimental and calculated data, respectively. The space group is cubic $Pn$-3 (201), with $a$=9.4947 \AA. $R_{\rm{wp}}$=0.1399, $R_{\rm{p}}$=0.1047, $\chi^2$=2.107. (c) Crystal structure of La$_3$Ir$_3$O$_{11}$. Grey, yellow and red spheres represent La, Ir and O atoms, respectively. (d) Schematic of Ir$_2$O$_{10}$ dimer. Different Ir-O distances indicate highly distorted octahedra. \label{f1}}
\end{figure}

The as-grown La$_3$Ir$_3$O$_{11}$ single crystals are black and exhibit cubic morphology, with a typical size of 0.5$\times$0.5$\times$0.5 mm$^3$, as seen in the scanning electron microscopy (SEM) image in Fig. \ref{f1}(a). Figure \ref{f1}(b) shows the powder XRD pattern, and the refined crystal structure is in good agreement with the $Pn$-3 (201) space group, like La$_3$Ru$_3$O$_{11}$, Bi$_3$Ru$_3$O$_{11}$ and Bi$_3$Os$_3$O$_{11}$ \cite{24}. As illustrated in Fig. \ref{f1}(c), the unit cell of La$_3$Ir$_3$O$_{11}$ contains six Ir$_2$O$_{10}$ dimers that are composed of pairs of edge-shared IrO$_6$ octahedra. The neighboring Ir ions form triangles which are located in perpendicular planes. From Fig. \ref{f1}(c), we can see that such triangular structures are connected in 3D space. Thus, the crystal structure of La$_3$Ir$_3$O$_{11}$ is geometrically frustrated, like the well-known pyrochlore structure, in which Ir ions form corner-shared tetrahedra in three dimensions \cite{Millican2007}. Such 3D geometrical frustration would cause strong magnetic frustration if we consider magnetic configurations. Even for a ferromagnetic (FM) arrangement, the magnetic structure can be still frustrated, due to the orthogonal distribution of $d$ orbitals. For an antiferromagnetic (AFM) arrangement, magnetic order cannot exist in the triangular structure.

When we zoom into a single Ir$_2$O$_{10}$ dimer [Fig. \ref{f1}(d)], we find that the octahedra are highly distorted, as evidenced by the different Ir-O distances and the unique Ir-O1-Ir bond angle (Table SI in Supplemental Material \cite{25}). Such a Ir-O1-Ir bond angle ($\sim$97.4$^\circ$) is much smaller than that of Sr$_2$IrO$_4$ (157$^\circ$) \cite{Huang1994} but close to that of Na$_2$IrO$_3$ (98$^\circ$) \cite{Choi2012}. Na$_2$IrO$_3$ possesses similar edge-shared IrO$_6$ octahedra.

The chemical stoichiometry is confirmed by the EDS characterization (Fig. S1 in Supplemental Material \cite{25}). According to the formula, the valence state of Ir ions is +4.33, an FVS between +4 and +5. Instead of an average result of mixed +4 and +5 valence states, the +4.33 valence state is statistically uniform, supported by previous M\"{o}ssbauer spectroscopy \cite{9}. This is reasonable if we notice that only one equivalent site of Ir is present in the crystal structure (Table SI in Supplemental Material \cite{25}). Imaging that the valence state is mixed +4 and +5, the crystal structure should be different, in view of the coupling between charge and lattice. Such an FVS is a result of prominent electron hopping within and between dimers. The enhanced hopping, together with the covalent bonding in the dimerized structure \cite{19}, increases the difficulty to determine the moment of Ir ions. Also, the magnetic structure of Ir network could be complicated due to the interplay of intra-dimer interactions, inter-dimer interactions and geometrical frustration.

\subsection{Magnetic, specific heat, and electrical measurements}

\begin{figure}[htbp]
\center
\includegraphics[width=8.5cm]{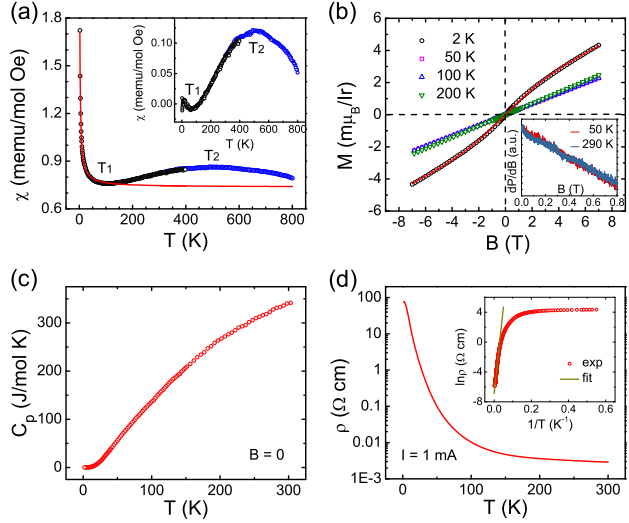}
\caption{(a) Temperature dependent susceptibility taken along [001] direction in a field of 0.1 T from 2 K to 400 K (black circles) and taken at 1 T from 300 K to 800 K (blue circles). Solid curve represents the fit to the Curie-Weiss law, $\chi_{\rm CW}=\chi_0+\frac{C}{T-\Theta}$, for the low-temperature data. Inset: the residual part after subtracting the Curie-Weiss part from the total susceptibility. (b) $M$-$H$ curves taken at various temperatures within $\pm$7 T. Solid curve is the fit to the Brillouin function for the 2 K data. Inset: ESR spectra taken at 50 K and 290 K with $B$$\parallel$[001]. (c) Heat capacity taken in the range of 2-300 K without a magnetic field. (d) Resistivity as a function of temperature measured with $I$=1 mA. Inset: the data between 34 K and 143 K fitted to the thermal activation model $\rho$($T$)=$\rho_0$exp($E_{\rm a}/k_{\rm B} T$). \label{f2}}
\end{figure}

Magnetization measurements were performed from 2 K to 800 K. As shown in Fig. \ref{f2}(a), with the ramping temperature, the susceptibility ($\chi$) first decreases sharply at low temperatures, and then increases almost linearly, leaving a minimum at $T_1$$\sim$70 K. The rising susceptibility persists up to $T_2$$\sim$500 K, above which a drop follows. In the Supplemental Material \cite{25}, we show that the ZFC (zero field cooling) and FC (field cooling) curves reveal no bifurcation and the anisotropy is also negligibly small.

As follows, we discuss the low-temperature and high-temperature data separately and in detail. The low-temperature $\chi$($T$) can be fitted to the Curie-Weiss law, i.e., $\chi_{\rm CW}=\chi_0+\frac{C}{T-\Theta}$, where $\chi_0$, $C$ and $\Theta$ are the temperature independent term, the Curie constant and the Weiss constant, respectively. It is noteworthy that the temperature range used for the fit is 2-50 K. Other ranges (2-35 K or 2-70 K) lead to almost the same results. $\chi_0$ is obtained as 7.37$\times$10$^{-4}$ emu/mol Oe. The effective moment $\mu_{\rm eff}$ is calculated to be 0.086 $\mu_{\rm B}$/Ir. $\Theta$=-0.91 K suggests very weak exchange interactions. In order to check the magnetic nature at low temperature, the magnetic field dependence ($M$-$H$) was measured. As seen in Fig. \ref{f2}(b), all the curves pass through the original point. While the 50 K, 100 K and 200 K data show a linear dependence, the 2 K data resemble a Brillouin-type curve \cite{28}, which can be formulated as $M=NgJ\mu_{\rm B}B_{\rm J}(gJ\mu_{\rm B}H/k_{\rm B} T)$. Here $B_{\rm J}(x)=\frac{2J+1}{2J}\coth\frac{(2J+1)x}{2J}-\frac{1}{2J}\coth\frac{x}{2J}$ is the Brillouin function \cite{29}, $N$ is the number of moments, $g$ is the Land\'{e} $g$ factor, and $J$ is the total angular momentum quantum number. By assuming $J$=1/2 (i.e., for Ir$^{4+}$ ions), a very small $N$=0.0012$N_{\rm Ir}$ ($N_{\rm Ir}$ is the number of all Ir ions) is obtained, which means that at 2 K only about 0.1\% of Ir ions contribute to the Curie-Weiss susceptibility. $\chi_{\rm CW}$ possibly originates from a small proportion of ``isolated'' Ir ions that are produced by the unavoidable defects in samples. That is to say, the low-temperature behavior is not intrinsic.

Hence the Curie-Weiss part should be subtracted from the total susceptibility when we discuss the high-temperature data. The inset of Fig. \ref{f2}(a) shows the residual susceptibility, labeled $\chi_{\rm NCW}$. There are three distinct features. First, the near-linear temperature dependence, between $T_1$ and $T_2$, deviates from the Curie-Weiss law. Second, a drop appears above $T_2$. Third, at zero temperature, the susceptibility approaches zero. However, it should be a finite value if we note that the $\chi_0$ obtained from the Curie-Weiss fit is relatively large, which possibly includes the part of $\chi_{\rm NCW}$($T$=0). When we examine these features, we should always keep in mind that the system is in a paramagnetic state. Such an uncommon non-Curie-Weiss paramagnetic susceptibility has not been reported for other iridates. The $\chi_{\rm NCW}$ indeed reflects the intrinsic magnetic property of La$_3$Ir$_3$O$_{11}$. It cannot be explained by the simple Curie-Weiss law, where only one kind of exchange interaction is assumed.

Moreover, the absence of magnetic hysteresis in $M$-$H$ curves suggests very weak or even no FM interactions, which is also confirmed by the ESR spectra. Typical ferromagnetic resonance (FMR) in iridates usually occurs around 0.2$\sim$0.4 T \cite{6,Bahr14}. As shown in the inset of Fig. \ref{f2}(b), such a mode is not observed up to 0.8 T. The ``smooth'' background looks like AFM or nonmagnetic materials. Apparently, La$_3$Ir$_3$O$_{11}$ is magnetic. Thus, the nature of magnetic interactions should be dominantly AFM.

Figure \ref{f2}(c) presents the specific heat data taken between 2 K and 300 K for La$_3$Ir$_3$O$_{11}$. No anomaly is observed in this temperature range, confirming the absence of long range magnetic order. Figure \ref{f2}(d) shows the temperature dependence of electrical resistivity, exhibiting a typical semiconductor behavior. Unlike other 4$d$ and 5$d$ isostructural TMOs that are usually metallic in transport \cite{15,24}, La$_3$Ir$_3$O$_{11}$ is an exception whose transport behavior violates the conventional band-driven itinerant picture. Besides, none of transport models can fit the resistivity data well for the whole temperature range. Similar phenomenon has been reported for other iridates \cite{Han2016}, due to the strong fluctuation arising from the enhanced SOC and the onsite Coulomb repulsion of 5$d$ electrons. The energy gap $\Delta$ ($\Delta$$\sim$2$E_{\rm a}$) can be estimated by piecewise fitting to the thermal activation model, i.e., $\rho$($T$)=$\rho_0$exp($E_{\rm a}/k_{\rm B} T$), where $E_{\rm a}$ is the activation energy and $k_{\rm B}$ is the Boltzmann constant. An energy gap of about 39 meV is obtained for the temperature range of 34-143 K [inset of Fig. \ref{f2}(d)].

\subsection{DFT calculations}

\begin{figure}[htbp]
\center
\includegraphics[width=8.5cm]{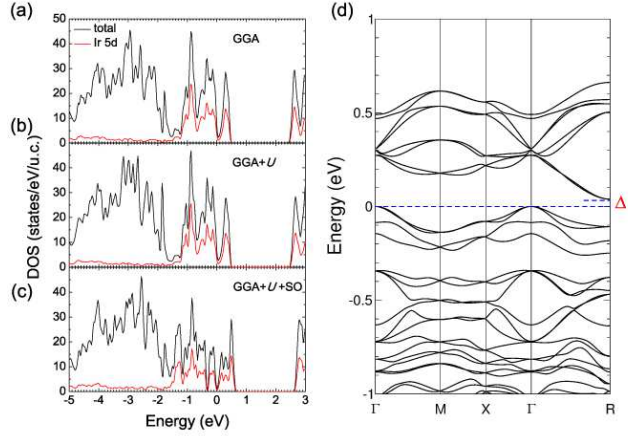}
\caption{DFT calculation results for La$_3$Ir$_3$O$_{11}$. (a)-(c) Total DOS (black) and the partial DOS for Ir 5$d$ orbitals (red) calculated with GGA, GGA+$U$, and GGA+$U$+SO ($U$=1.5 eV), respectively. (d) Band structures calculated with GGA+$U$+SO ($U$=1.5 eV). An indirect gap forms between the top of valence band (i.e., $\Gamma$ point) and the bottom of conduction band (i.e., R point). \label{f3}}
\end{figure}

To understand the unconventional insulating transport, we further performed DFT calculations for La$_3$Ir$_3$O$_{11}$. A simple but sufficiently effective method, i.e., GGA+$U$+SO, was adopted to calculate the density of states (DOS) and band structures. As shown in Figs. \ref{f3}(a)-\ref{f3}(c), the total DOS (black color) and the partial DOS for Ir 5$d$ orbitals (red color) are plotted as a function of energy for GGA, GGA+$U$, and GGA+$U$+SO ($U$=1.5 eV), respectively. The big gap ($\sim$2 eV), between 0.5 eV and 2.5 eV, should be attributed to the octahedral crystal field that splits the $e_{\rm g}$ and $t_{\rm 2g}$ orbitals \cite{5}. Comparing Figs. \ref{f3}(a) and \ref{f3}(b), changes can be hardly found when $U$ is applied for the 5$d$ orbitals of Ir. Remarkable changes occur once SO is further included into calculations. The $t_{\rm 2g}$ orbitals are split by the SOC, and more importantly, a small gap appears at the Fermi level [Fig. \ref{f3}(c)]. To elucidate the details around the Fermi level, the band structures are also plotted in Fig. \ref{f3}(d). The top of valence band is located at the $\Gamma$ point of Brillouin zone, and the bottom of conduction band is located at the R point, forming an indirect gap of $\sim$29 meV. This is consistent with the fitting result of resistivity data.

By tuning parameters, the origin of the gap can be revealed. Table SII in Supplemental Material summarizes the calculation results with different settings of parameters \cite{25}. It is found that GGA solely, or only one correction term ($U$ or SO) added, would not lead to an insulating gap. The gap can be created only when $U$ and SO are both considered. We also note that as $U$ is increased from 1.5 eV to 3 eV, the gap is enlarged to 62 meV. Therefore, the gap is a cooperative result of $U$ and SO interactions.

\begin{figure}[htbp]
\center
\includegraphics[width=8.5cm]{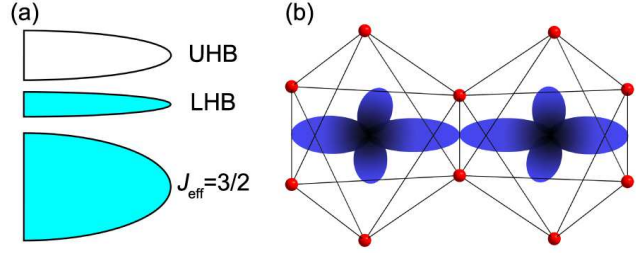}
\caption{(a) Energy diagram for the 5$d^{4.67}$ ($t_{\rm 2g}^{4.67}$) configuration. (b) Schematic $d_{xy}$ orbital in an edge-shared dimer, with large overlap. \label{f4}}
\end{figure}

Based on the calculations, we illustrate the energy diagram for the 5$d^{4.67}$ ($t_{\rm 2g}^{4.67}$) electron configuration [Fig. \ref{f4}(a)]. The energetically lower $J_{\rm eff}$=3/2 band is filled by four electrons, i.e., fully filled, while the higher $J_{\rm eff}$=1/2 band is split into a lower Hubbard band (LHB, filled) and an upper Hubbard band (UHB, empty). This picture is generally consistent with that in Sr$_2$IrO$_4$, and well explains the insulating transport of La$_3$Ir$_3$O$_{11}$. The difference is the valence state of Ir ions, which is +4 in Sr$_2$IrO$_4$ and is fractional (+4.33) here. The occupancy of LHB is thus different. For Sr$_2$IrO$_4$, the fifth electron occupies the LHB, i.e., the unpaired $J_{\rm eff}$=1/2 state. This is the origin of magnetic moment of Ir$^{4+}$, and the inter-site hopping provides conductivity. But for La$_3$Ir$_3$O$_{11}$, the partial occupation (0.67) in LHB means existence of holes, allowing more inter-site hoppings than Sr$_2$IrO$_4$. The hopping can be further enhanced by the dimerized structure. As depicted in Fig. \ref{f4}(b), the overlapped $d_{xy}$ orbitals in an edge-shared geometry can serve as the tunnel of electron hoppings \cite{19}. Since the strong hopping will promote uniform charge distribution on all Ir ions, it is consistent with the single valence state of Ir ions claimed in Ref. \cite{9}. We may note that the $J_{\rm eff}$=1/2 picture still holds in La$_3$Ir$_3$O$_{11}$, in spite of the distortions of IrO$_6$ octahedra [Fig. \ref{f1}(d)]. As mentioned above, La$_3$Ir$_3$O$_{11}$ and Na$_2$IrO$_3$ possess similar edge-shared IrO$_6$ octahedra and a close Ir-O1-Ir bond angle. For Na$_2$IrO$_3$, it has been demonstrated that, though there exists a sizable trigonal distortion, the induced trigonal crystal field is negligibly small. As a consequence, the SOC-related $J_{\rm eff}$=1/2 scenario is still valid \cite{Gretarsson2013}.

Although the insulating transport is supported by the calculations, the obtained moments of Ir ions are unsatisfactory. For GGA or GGA+$U$, the calculated moment is about 0.16 $\mu_{\rm B}$/Ir, smaller than that of Sr$_2$IrO$_4$ (0.36 $\mu_{\rm B}$/Ir). This is reasonable because in La$_3$Ir$_3$O$_{11}$ the number of localized electrons is less and the enhanced hopping will further reduce the moment. Nevertheless, the calculation seems to fail once SO is considered: the calculated moments (total, spin, and orbital moments) become nearly zero (see Table SII in Supplemental Material \cite{25}). This phenomenon is rarely seen, as the SO interactions will not eliminate the moment but just reduce it slightly \cite{19}. The GGA result (0.16 $\mu_{\rm B}$/Ir) provides an important reference for the following discussion, i.e., the upper limit of Ir moments.

\subsection{Theoretical model of susceptibility}

\begin{figure}[htbp]
\center
\includegraphics[width=8.5cm]{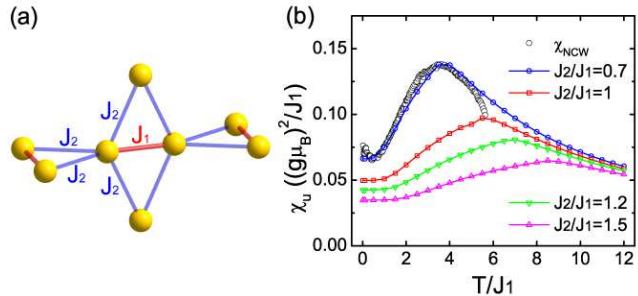}
\caption{(a) Schematic of multiple interactions in La$_3$Ir$_3$O$_{11}$. Yellow spheres represent Ir ions; red and blue lines represent the intra-dimer ($J_1$, nearest-neighbor) and inter-dimer ($J_2$, next-nearest-neighbor) interactions, respectively. (b) Calculated temperature dependence of $\chi_{\rm u}$ with different scales of $J_2$/$J_1$. Black circles represent the experimental data ($\chi_{\rm NCW}$) that most fit the $J_2$/$J_1$=0.7 curve. \label{f5}}
\end{figure}

To understand the non-Curie-Weiss paramagnetism, we attempt to disclose the underlying physical origin. Although the high temperature susceptibility hump looks like low dimensional or dimer system, we exclude the possibilities after analyses. Similar behavior has been observed in some one-dimensional and two-dimensional Heisenberg spin-1/2 systems with a spin gap in their excitation spectra, e.g., SrCu$_2$O$_3$ \cite{Azuma1994}. However, this possibility can be ruled out in consideration of the completely 3D character of the Ir network in La$_3$Ir$_3$O$_{11}$. Another possible origin is electronic dimerization which often leads to a dimer-like spin configuration and opens a spin gap via spin-Peierls transition. The transition is usually accompanied by a lattice distortion \cite{Kuroe1994,Fischer1998} or a sharp drop in susceptibility \cite{Castilla1995,Hemberger1998}. However, such a transition is not observed in La$_3$Ir$_3$O$_{11}$. As shown in Fig. S2 \cite{25}, the powder XRD taken at various temperatures (from 35 K to 300 K) shows no signature of lattice distortion. Also, the susceptibility is a broad peak instead of a sharp drop. There is no direct evidence for electronic dimerization in current experimental data.

Here we perform modeling and simulation work in the regime of competing nearest-neighbor (NN) and next-nearest-neighbor (NNN) interactions. As illustrated in Fig. \ref{f5}(a), for each Ir ion, there is one (intra-dimer) NN Ir ion and four (inter-dimer) NNN Ir ions. The NN and NNN interactions are labeled $J_1$ and $J_2$, respectively. Four $J_2$ interactions are located in two perpendicular planes, and within each plane the $J_1$ and $J_2$ interactions form a triangle, both in a sign of strong frustration. In the present paper, we start from a quantum spin model and use finite temperature statistical mechanics \cite{30,31}. For La$_3$Ir$_3$O$_{11}$, the following frustrated $J_1$-$J_2$ Heisenberg model is assumed,
\begin{eqnarray}
\mathcal{H}=J_{1}\sum_{{\langle}i,j{\rangle}}\mathbf{S}_i{\cdot}\mathbf{S}_j+J_{2}\sum_{{\langle}{\langle}i,j{\rangle}{\rangle}}\mathbf{S}_i{\cdot}\mathbf{S}_j,
\end{eqnarray}
where ${\langle}i,j{\rangle}$ and ${\langle}{\langle}i,j{\rangle}{\rangle}$ denote the summations over the nearest and next-nearest neighbors, respectively. After transformation, diagonalization and other mathematic treatments (detailed process can be found in part 2 of Supplemental Material \cite{25}), the ground state energy $\epsilon_{\mathbf{k}}$ is deduced, and then the static uniform magnetic susceptibility is derived as
\begin{eqnarray}
\chi_{\rm u}=\frac{(g\mu_{\rm B})^{2}}{4k_{\rm B}TN}\sum_{\mathbf{k}}\frac{1}{\sinh^2\left(\frac{\epsilon_{\mathbf{k}}}{2k_{\rm B}T}\right)} \label{Eq:SuspA}.
\end{eqnarray}

Equation (\ref{Eq:SuspA}) can be solved at finite temperatures through a numerical calculation process (see part 3 of Supplemental Material for details \cite{25}). Figure \ref{f5}(b) shows the temperature dependence of $\chi_{\rm u}$ with different energy scales of $J_2$/$J_1$. Interestingly, all curves show an increase in susceptibility with the ramping temperature for a wide range of $T$/$J_1$, then followed by a drop above a comparably high temperature. For a larger $J_2$/$J_1$, the susceptibility rises slowly but persists to a higher temperature. Just as important, the curves approach a finite value at zero temperature. Hence the new theoretical model reproduces all the three features of the non-Curie-Weiss susceptibility as described in Fig. \ref{f2}(a), showing high consistence between our theory and experiment.

Detailed comparison to experiment can reveal more useful information. Given the increase of susceptibility from $T_1$ to $T_2$ (i.e., $\Delta\chi_{\rm NCW}$=1.2$\times$10$^{-4}$ emu/mol Oe for $T_2$/$T_1$$\sim$7), we find that the situation in La$_3$Ir$_3$O$_{11}$ coincides most with $J_2$/$J_1$$\sim$0.7 (see black circles and blue curve in Fig. \ref{f5}(b)). In such a case, $J_1$ is estimated to be 12 meV, which is a reasonable energy scale for the dimerized iridates \cite{10}. The ``effective" $g$ factor is obtained as 0.8, suggesting a considerably small effective moment of Ir ions. Moreover, the susceptibility at zero temperature is about 1.1$\times$10$^{-4}$ emu/mol Oe, smaller than $\chi_0$ that is obtained in the Curie-Weiss fit (i.e., 7.37$\times$10$^{-4}$ emu/mol Oe). This is necessary because $\chi_0$ includes other temperature independent contribution, such as Van Vleck paramagnetism \cite{32}. By further comparing $\chi_{\rm u}$($T_1$) with $\frac{N\mu_{\rm eff}^2}{3k_{\rm B}T_{\rm 1}}$, we give a rough estimate for the spin moment of Ir ions, i.e., 0.146 $\mu_{\rm B}$. This value is slightly smaller than the calculated moment obtained from the DFT calculations (0.16 $\mu_{\rm B}$) and is consistent with the small ``effective" $g$ factor. From these analyses, we can deduce that in La$_3$Ir$_3$O$_{11}$ the frustration induced non-Curie-Weiss paramagnetism forbids a long-range magnetic order, even at zero temperature. This might imply a possible QSL state that is pursued hard recently \cite{33}.

Last, we talk about possible reasons for the reduced moment of Ir ions. For an unpaired electron occupying the $J_{\rm eff}$=1/2 level as stated in Fig. \ref{f4}(a), the expected ionic moment is 1 $\mu_{\rm B}$/Ir. This value will be decreased to 0.67 $\mu_{\rm B}$/Ir, taking into account the partial occupation in La$_3$Ir$_3$O$_{11}$. However, the actual moment is much smaller ($<$ 0.16 $\mu_{\rm B}$). Besides oxygen covalency, some other origins that are unique in La$_3$Ir$_3$O$_{11}$ could be responsible for the reduction. One is the enhanced $d$-$d$ hopping between neighboring Ir ions. Once a partial occupancy is existing in the $J_{\rm eff}$=1/2 level, the emerging possibility of empty states (i.e., holes) allows more hoppings between the empty states and occupied states. As a result, the itinerancy of electrons is increased, which reduces the localized moment. Another origin is the strong orbital-selective covalent bonding facilitated by the dimerized structure \cite{19}. In such a geometry, the $d_{xy}$ orbitals have large overlap and form inter-site covalent bonds [Fig. \ref{f4}(b)], on which the electrons are nonmagnetic. For an iridium dimer, SOC usually works together with covalent bonding and further reduces the moment \cite{11}. In one word, the greatly reduced moment in La$_3$Ir$_3$O$_{11}$ can be understood in terms of the combined effect of enhanced electron hopping, covalent bonding, and SOC.

Although the exact moment of Ir is not experimentally determined in the present study, we propose possible methods for future experiments. More efforts are worth it as the eventual solution of this problem will help understand the nature of magnetic interactions and ground state in La$_3$Ir$_3$O$_{11}$. One method is performing the Curie-Weiss fit at very high temperatures ($>$800 K). When $T\gg J_1$, $J_2$, the Curie-Weiss behavior is restored. Besides, magnetic X-ray absorption spectroscopy is another feasible technique.

\section{conclusions}

Itinerancy and localization are at the very heart of magnetism and other properties of TMOs. Since an insulating gap is present in La$_3$Ir$_3$O$_{11}$, the energetically active electrons are more localized. This may explain why the localized Heisenberg model can well reproduce the non-Curie-Weiss behavior. However, the reduced moment reflects a certain degree of itinerancy in this system, where the enhanced hopping and inter-site covalent bonding cannot be underestimated. Such a situation is different from that in the similar compound La$_4$Ru$_6$O$_{19}$ \cite{15}, where the transport is metallic and conduction electrons interact with localized electrons. This Kondo mechanism leads to a non-Fermi-liquid behavior, as in other ruthenates \cite{34} and heavy fermion intermetallics \cite{35}. Thus, from the aspect of itinerancy and localization, La$_3$Ir$_3$O$_{11}$ is close to the localization limit, and the strong frustration and competing interactions prohibit a magnetic order. This suggests a possible candidate for the QSL state, although further conclusive work is needed.

\begin{acknowledgments}
We highly appreciate the insightful discussions with Prof. Gang Chen. This work was supported by the National Key R\&D Program of China (Grant Nos. 2016YFA0300404, 2017YFA0403600 and 2017YFA0403502), and the National Natural Science Foundation of China (Grant Nos. U1532267, 11674327, U1932216, 11874363, 51603207, 11574288 and U1732273). J.Y., J.R.W., and W.L.Z. contributed equally to this work.
\end{acknowledgments}

\end{document}


\pagestyle{plain}

\begin{center}
\textbf{Supplementary Information for ``Frustration induced non-Curie-Weiss paramagnetism in La$_3$Ir$_3$O$_{11}$: a fractional-valence-state iridate''}
\end{center}

\vspace{2ex}
\noindent {\textbf{1. Miscellaneous results}}

\begin{table}[hbp]
\begin{threeparttable}
\caption{Refined structural parameters of La$_3$Ir$_3$O$_{11}$.}\label{tab1}
\begin{tabular}{llllll|lll}
\hline
\hline
Atom&Wyck.&$x$&$y$&$z$&U$_{\rm{iso}}$&$d_{\rm{Ir-O1}}$ ({\AA})&$d_{\rm{Ir-O3}}$ ({\AA})&$d_{\rm{Ir-Ir}}$ ({\AA})\\
\hline
La1&4$b$&0&0&0&0.0083&1.9950($\times$2)&1.9777($\times$2)&2.9975\\
La2&8$e$&0.3860&0.3860&0.3860&0.0090&&2.0156($\times$2)&3.5776($\times$4)\\
Ir&12$g$&0.4079&0.75&0.25&0.0011&&&\\
O1&12$f$&0.6113&0.25&0.25&0.0650&$\varphi_{\rm{Ir-O1-Ir}}$ ($^\circ$)&$\varphi_{\rm{Ir-O3-Ir}}$ ($^\circ$)&\\
\cline{7-9}
O2&8$e$&0.1443&0.1443&0.1443&0.0253&97.402&127.246&\\
O3&24$h$&0.5854&0.2528&0.5418&0.0285&&&\\
\hline
\hline
\end{tabular}
\begin{tablenotes}
\item[]The space group is cubic $Pn$-3 (201), with $a$=9.4947 \AA. $R_{\rm{wp}}$=0.1399, $R_{\rm{p}}$=0.1047, $\chi^2$=2.107.
\end{tablenotes}
\end{threeparttable}
\end{table}

\begin{table}[htbp]
\caption{Calculation results with GGA+$U$+SO for La$_3$Ir$_3$O$_{11}$.}\label{tab2}
\begin{tabular}{lccccccc}
\hline
\hline
&GGA&\multicolumn{2}{c}{GGA+$U$}&GGA+SO&\multicolumn{3}{c}{GGA+$U$+SO}\\
&&$U$=1.5 eV&$U$=2 eV&&$U$=1.5 eV&$U$=2 eV&$U$=3 eV\\
\hline
Gap (meV)&0&0&0&0&29&40&62\\
MMI ($\mu_{\rm{B}}$)&0.1579&0.1592&0.1593&-0.00005&-0.00008&-0.00019&-0.00040\\
&&&&0.00003&0.00006&0.00011&0.00005\\
&&&&0.00001&0.00001&0.00003&0.00011\\
SPI$_z$ ($\mu_{\rm{B}}$)&-&-&-&-0.00001&0.00000&0.00000&-0.00041\\
&&&&0.00000&0.00001&0.00000&0.00000\\
&&&&0.00000&0.00000&0.00000&0.00001\\
ORB$_z$ ($\mu_{\rm{B}}$)&-&-&-&0.00000&0.00000&0.00000&-0.00040\\
&&&&0.00000&0.00000&0.00000&-0.00028\\
&&&&0.00000&0.00000&0.00000&0.00000\\
\hline
\hline
\end{tabular}
\end{table}

In the spin-polarized DFT calculations, the magnetic configuration was set as ferromagnetism, and the magnetic moment was along $z$ axis for the calculations with SO. MMI, SPI$_z$, and ORB$_z$ in Table S\ref{tab2} represent magnetic moment, $z$ component of spin moment, and $z$ component of orbital moment, respectively.

\begin{figure}[htbp]
\center
\includegraphics[width=3.5in]{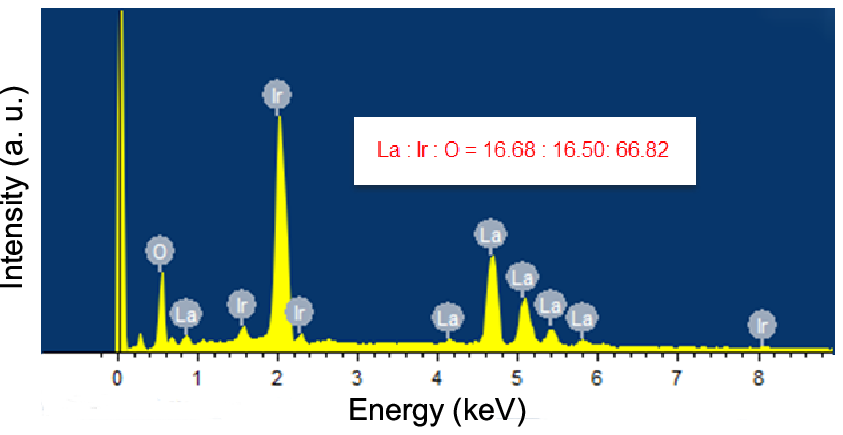}
\caption{EDS result of single crystal La$_3$Ir$_3$O$_{11}$. The ratio of elements is generally consistent with the formula stoichiometry.}
\end{figure}

\begin{figure}[htbp]
\center
\includegraphics[width=3.1in]{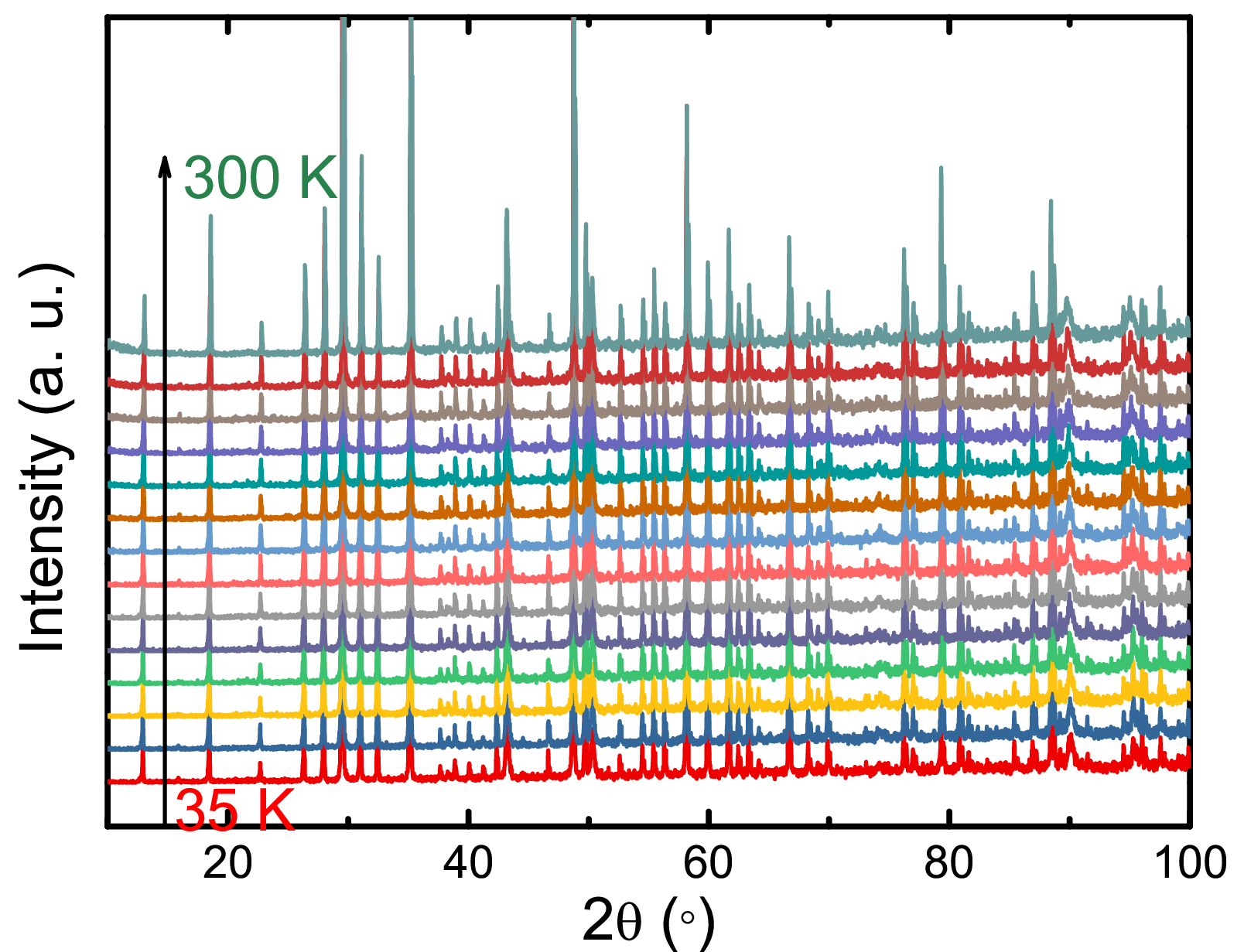}
\caption{Powder XRD patterns of La$_3$Ir$_3$O$_{11}$ taken at various temperatures from 300 K down to 35 K. No signature of lattice distortion is observed.}
\end{figure}

\begin{figure}[htbp]
\center
\includegraphics[width=6.2in]{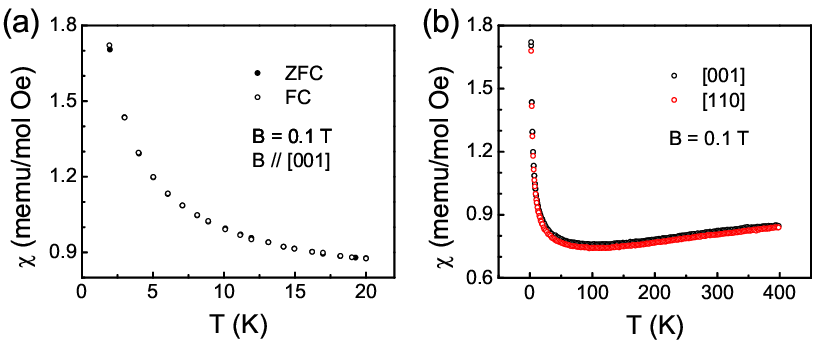}
\caption{Temperature dependence of susceptibility taken at a field of 0.1 T for La$_3$Ir$_3$O$_{11}$. (a) ZFC and FC curves; (b) along [001] and [110] directions.}
\end{figure}

\newpage
\noindent {\textbf{2. Theoretical model}}

In the present paper, we start from a quantum spin model and use finite temperature statistical mechanics \cite{AuerbachBook, ZhangGuangMing09}. For La$_3$Ir$_3$O$_{11}$, the following frustrated antiferromagnetic $J_1$-$J_2$ Heisenberg model is assumed,
\begin{eqnarray}
\mathcal{H}=J_{1}\sum_{{\langle}i,j{\rangle}}\mathbf{S}_i{\cdot}\mathbf{S}_j+J_{2}\sum_{{\langle}{\langle}i,j{\rangle}{\rangle}}\mathbf{S}_i{\cdot}\mathbf{S}_j,
\end{eqnarray}
where ${\langle}i,j{\rangle}$ and ${\langle}{\langle}i,j{\rangle}{\rangle}$ denote the summations over the nearest and next-nearest neighbors, respectively, which are defined in Fig. 5(a) in the main text. According to two lattice model, the lattice is divided into A and B sublattices. On the A (B) lattice, the vacuum state is the $S^z=S$ (-$S$) state. There are two spins in each unit state.

We then use the antiferromagnetic Dyson-Maleev transformation to represent the spin operators:
\begin{eqnarray}
&S&_l^-=a_l^\dag,~~~~S_l^+=(2S-a_l^{\dag}a_l)a_l,\nonumber
\\
&S&_l^z=S-a_l^{\dag}a_l,~~~~l{\in}A,\nonumber
\\
&S&_m^-=b_m,~~~~S_m^+=b_m^{\dag}(2S-b_m^{\dag}b_m),\nonumber
\\
&S&_m^z=-S+b_m^{\dag}b_m,~~~~m{\in}B,
\end{eqnarray}
where $a_l^\dag$, $a_l$, $b_m^\dag$, and $b_m$ are bosonic operators. After Fourier transformation, the model becomes
\begin{eqnarray}
\mathcal{H}{\approx}\sum_{\mathbf{k}}[{\eta}_\mathbf{k}(a_\mathbf{k}^{\dag}a_\mathbf{k}+b_\mathbf{k}^{\dag}b_\mathbf{k})+
{\Lambda}_\mathbf{k}(a_\mathbf{k}b_{-\mathbf{k}}+a_\mathbf{k}^{\dag}b_{-\mathbf{k}}^\dag)]-2NS(J_2S+\lambda),
\end{eqnarray}
where
\begin{eqnarray}
\eta_{\mathbf{k}}&=&2J_{1}S\cos(2k_{x})+2J_{2}S+\lambda,
\\
\Lambda_{\mathbf{k}}&=&2J_{1}S+2J_{2}S\gamma_{\mathbf{k}},
\end{eqnarray}
with
\begin{eqnarray}
\gamma_{\mathbf{k}}&=&4\cos(2k_{x})\sin(k_{y})+4\cos(k_{x})\sin(2k_{z}).
\end{eqnarray}
A chemical potential term $\lambda$ is introduced to make the local magnetization vanish at finite temperatures.

By using the Bogoliubov transformation, the Hamiltonian is diagonalized as
\begin{eqnarray}
\mathcal{H}=\sum_{\mathbf{k}}\epsilon_\mathbf{k}(\alpha_\mathbf{k}^{\dag}\alpha_\mathbf{k}+\beta_{-\mathbf{k}}^{\dag}\beta_{-\mathbf{k}})+\varepsilon_gN,
\end{eqnarray}
where
\begin{eqnarray}
\epsilon_{\mathbf{k}}&=&\sqrt{\eta_{\mathbf{k}}^{2}-\Lambda_{\mathbf{k}}^{2}},
\end{eqnarray}
and $\varepsilon_g$ is the ground state energy per site. $\lambda$ is determined by the condition
\begin{eqnarray}
\frac{1}{N}\sum_{\mathbf{k}}\frac{\eta_{\mathbf{k}}}{\epsilon_{\mathbf{k}}}\coth\left(\frac{\epsilon_{\mathbf{k}}}{2k_{B}T}\right)=2S+1. \label{Eq:ConstraintA}
\end{eqnarray}

Based on the above results, the static uniform susceptibility is derived as \cite{AuerbachBook, ZhangGuangMing09}
\begin{eqnarray}
\chi_{u}=\frac{(g\mu_{B})^{2}}{4k_{B}TN}\sum_{\mathbf{k}}\frac{1}{\sinh^2\left(\frac{\epsilon_{\mathbf{k}}}{2k_{B}T}\right)} \label{Eq:SuspA},
\end{eqnarray}
which needs to be multiplied by a factor of 1/3, compared to the expression obtained from spin-correlation function.\\

\noindent {\textbf{3. Numerical calculation of susceptibility}}

Utilizing the replacement \cite{AuerbachBook}
\begin{eqnarray}
\frac{1}{N}\sum_{\mathbf{k}}\rightarrow\int\frac{d^3\mathbf{k}}{(2\pi)^{3}}\equiv\int_{-\pi}^{\pi}\frac{dk_{x}}{2\pi}\int_{-\pi}^{\pi}
\frac{dk_{y}}{2\pi}\int_{-\pi}^{\pi}\frac{dk_{z}}{2\pi},
\end{eqnarray}
equation~(\ref{Eq:ConstraintA}) becomes
\begin{eqnarray}
\int\frac{d^3\mathbf{k}}{(2\pi)^{3}}\frac{\eta_{\mathbf{k}}}{\epsilon_{\mathbf{k}}}\coth\left(\frac{\epsilon_{\mathbf{k}}}{2k_{B}T}\right)=2S+1.  \label{Eq:ConstraintB}
\end{eqnarray}
If we can get a solution for $\lambda$ from equation~(\ref{Eq:ConstraintB}), the system is in the magnetic disordered phase.
Then, we substitute the value of $\lambda$ into the following equation
\begin{eqnarray}
\chi_{u}=\frac{(g\mu_{B})^{2}}{4k_{B}T}\int\frac{d^3\mathbf{k}}{(2\pi)^{3}}\frac{1}{\sinh^2\left(\frac{\epsilon_{\mathbf{k}}}{2k_{B}T}\right)}, \label{Eq:SuspB}
\end{eqnarray}
and obtain the value of susceptibility directly.  If there is no solution for $\lambda$ from  equation~(\ref{Eq:ConstraintB}), equation~(\ref{Eq:ConstraintB}) should be replaced
with
\begin{eqnarray}
m_{0}+\left.\int\frac{d^3\mathbf{k}}{(2\pi)^{3}}\frac{\eta_{\mathbf{k}}}{\epsilon_{\mathbf{k}}}
\coth\left(\frac{\epsilon_{\mathbf{k}}}{2k_{B}T}\right)\right|_{\lambda=\lambda_{min}}=2S+1,  \label{Eq:ConstraintC}
\end{eqnarray}
where $m_{0}$ is corresponding to the magnitude of order parameter of magnetic ordered phase \cite{AuerbachBook}.
If $\lambda<\lambda_{min}$, the dispersion $\epsilon_{\mathbf{k}}$ becomes imaginary in some region of momentum space. In the numerical calculation, we
should adjust $\lambda$ so that $\epsilon_{\mathbf{k}}$ takes real value in the whole region of momentum space.
If $\lambda=\lambda_{min}$, $\epsilon_{\mathbf{k}}$ becomes gapless at some
points $\mathbf{k}_{0}^{i}$ of momentum space, and the bosons are condensed at these points. If $\lambda>\lambda_{min}$, $\epsilon_{\mathbf{k}}$  becomes gapped in the whole region
of momentum space. The parameter $m_{0}$ is given by
\begin{eqnarray}
m_{0}=\left.\sum_{i}\frac{1}{N}\frac{\eta_{\mathbf{k}_{0}^{i}}}{\epsilon_{\mathbf{k}_{0}^{i}}}\coth\left(\frac{\epsilon_{\mathbf{k}_{0}^{i}}}{2k_{B}T}\right)\right|_{\lambda\rightarrow\lambda^{min}+0^{+}}
=\left.\sum_{i}\frac{1}{N}\eta_{\mathbf{k}_{0}^{i}}\frac{2k_{B}T}{\epsilon_{\mathbf{k}_{0}^{i}}^{2}}\right|_{\lambda\rightarrow\lambda^{min}+0^{+}}. \label{Eq:m0DefA}
\end{eqnarray}
In current case, the susceptibility satisfies
\begin{eqnarray}
\chi_{u}=(g\mu_{B})^{2}\left[m_{0}'+\frac{1}{4k_{B}T}\int\frac{d^3\mathbf{k}}{(2\pi)^{3}}\left.\frac{1}{\sinh^2\left(\frac{\epsilon_{\mathbf{k}}}{2k_{B}T}\right)}\right|_{\lambda=\lambda_{min}}\right], \label{Eq:SuspC}
\end{eqnarray}
where
\begin{eqnarray}
m_{0}'=\left.\sum_{i}\frac{1}{4k_{B}T}\frac{1}{N}\frac{1}{\sinh^{2}\left(\frac{\epsilon_{\mathbf{k}_{0}^{i}}}{2k_{B}T}\right)}\right|_{\lambda\rightarrow\lambda_{min}+0^{+}}
=\left.\sum_{i}\frac{1}{N}\frac{k_{B}T}{\epsilon_{\mathbf{k}_{0}^{i}}^{2}}\right|_{\lambda\rightarrow\lambda_{min}+0^{+}}. \label{Eq:moprimeDefA}
\end{eqnarray}

\begin{figure}[htbp]
\center
\includegraphics[width=6.2in]{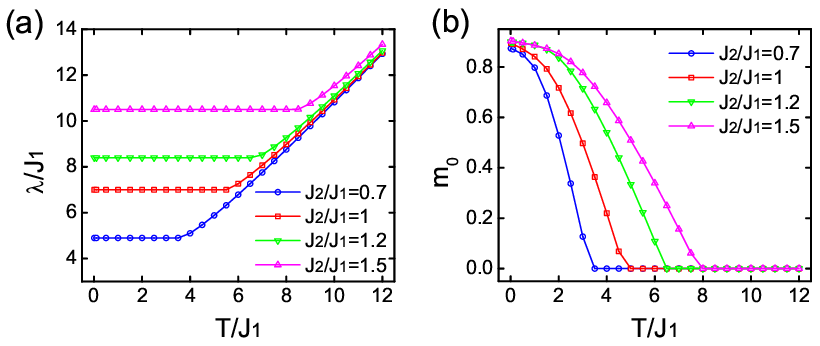}
\caption{Dependence of $\lambda$ and $m_{0}$ on temperature with different values of $J_{2}/J_{1}$. \label{Fig:lambdam0FinitT3D}}
\end{figure}

We find that in the case  $J_{2}/J_{1}>0.5$, if $\lambda=\lambda_{min}$ the dispersion $\epsilon_{\mathbf{k}}$ is gapless at the points
\begin{eqnarray}
&&\mathbf{k}_{0}^{1}=\left(0,\frac{\pi}{2}, \frac{\pi}{4},\right),
\quad\mathbf{k}_{0}^{2}=\left(0, \frac{\pi}{2},-\frac{3\pi}{4}\right),
\quad\mathbf{k}_{0}^{3}=\left(\pi, \frac{\pi}{2}, -\frac{\pi}{4}\right),\nonumber
\\
&&\mathbf{k}_{0}^{4}=\left(\pi, \frac{\pi}{2},\frac{3\pi}{4}\right),
\quad \mathbf{k}_{0}^{5}=\left(-\pi, \frac{\pi}{2}, -\frac{\pi}{4}\right),
\quad \mathbf{k}_{0}^{6}=\left(-\pi, \frac{\pi}{2}, \frac{3\pi}{4}\right).
\end{eqnarray}
Expanding the energy dispersion around these points, we obtain
\begin{eqnarray}
\lambda_{min}=14J_{2}S,
\end{eqnarray}
and find that the dispersion in the vicinity of these points can be approximated as
\begin{eqnarray}
\epsilon_{\mathbf{k}}
&=&\sqrt{v_{x}^{2}K_{x}^{2}+v_{y}^{2}K_{y}^{2}+v_{z}^{2}K_{z}^{2}},
\end{eqnarray}
where
\begin{eqnarray}
v_{x}=4S\sqrt{(5J_{2}-J_{1})\left(J_{1}+8J_{2}\right)},\quad
v_{y}=4S\sqrt{J_{2}\left(J_{1}+8J_{2}\right)},\quad
v_{z}=8S\sqrt{J_{2}\left(J_{1}+8J_{2}\right)}.
\end{eqnarray}
$\mathbf{K}$ measures the distance from some point $\mathbf{k}_{0}^{i}$.
For the case $J_{2}/J_{1}>0.5$, $m_{0}$ takes the form
\begin{eqnarray}
m_{0}
=\left.\sum_{i}\frac{1}{N}\left(2J_{1}S+2J_{2}S+\lambda_{min}\right)\frac{2k_{B}T}{\epsilon_{\mathbf{k}_{0}^{i}}^{2}}\right|_{\lambda\rightarrow\lambda^{min}+0^{+}}. \label{Eq:m0DefB}
\end{eqnarray}
From equations~(\ref{Eq:moprimeDefA}) and (\ref{Eq:m0DefB}), we get
\begin{eqnarray}
m_{0}'=\frac{1}{2}\frac{m_{0}}{2J_{1}S+2J_{2}S+\lambda_{min}}. \label{Eq:moprimeDefB}
\end{eqnarray}
Accordingly, the susceptibility can be further written as
\begin{eqnarray}
\chi_{u}=(g\mu_{B})^{2}\left[\frac{1}{2}\frac{m_{0}}{2J_{1}S+2J_{2}S+\lambda_{min}}
+\left.\frac{1}{4k_{B}T}\int\frac{d^3\mathbf{k}}{(2\pi)^{3}}\frac{1}{\sinh^2\left(\frac{\epsilon_{\mathbf{k}}}{2k_{B}T}\right)}\right|_{\lambda=\lambda_{min}}\right]. \label{Eq:SuspD}
\end{eqnarray}
After performing numerical calculation through equations~(\ref{Eq:ConstraintB}), (\ref{Eq:SuspB}), (\ref{Eq:ConstraintC}), (\ref{Eq:SuspC}), and (\ref{Eq:SuspD}),
we can obtain the solutions of parameters $\lambda$, $m_{0}$, and the susceptibility $\chi_{u}$. We present the temperature dependence of $\lambda$ and $m_{0}$
in Fig.~S\ref{Fig:lambdam0FinitT3D}, and the temperature dependence of $\chi_{u}$ in Fig. 5(b) in the main text
, for several different values of $J_{2}/J_{1}$.

